\documentclass[a4paper]{article}

\usepackage{INTERSPEECH2020}
\usepackage[colorlinks,
linkcolor=blue,
anchorcolor=green,
citecolor=green]{hyperref}
\usepackage{cite}
\usepackage{makecell}
\title{DPCRN: Dual-Path Convolution Recurrent Network for Single Channel Speech Enhancement}
\name{Xiaohuai Le$^1$$^,$$^2$$^,$$^3$, Hongsheng Chen$^1$$^,$$^2$$^,$$^3$, Kai Chen$^1$$^,$$^2$$^,$$^3$, Jing Lu$^1$$^,$$^2$$^,$$^3$}
\address{
  $^1$Key Laboratory of Modern Acoustics, Nanjing University, Nanjing 210093, China\\
  $^2$NJU-Horizon Intelligent Audio Lab, Horizon Robotics, Beijing 100094, China\\
  $^3$Nanjing Institute of Advanced Artificial Intelligence, Nanjing 210014, China}
\email{\{mg20220173, hschen\}@smail.nju.edu.cn, \{chenkai, lujing\}@nju.edu.cn}

\begin{document}
\maketitle
\begin{abstract}
The dual-path RNN (DPRNN) was proposed to more effectively model extremely long sequences for speech separation in the time domain. By splitting long sequences to smaller chunks and applying intra-chunk and inter-chunk RNNs, the DPRNN reached promising performance in speech separation with a limited model size. In this paper, we combine the DPRNN module with Convolution Recurrent Network (CRN) and design a model called Dual-Path Convolution Recurrent Network (DPCRN) for speech enhancement in the time-frequency domain. We replace the RNNs in the CRN with DPRNN modules, where the intra-chunk RNNs are used to model the spectrum pattern in a single frame and the inter-chunk RNNs are used to model the dependence between consecutive frames. With only 0.8M parameters, the submitted DPCRN model achieves an overall mean opinion score (MOS) of 3.57 in the wide band scenario track of the Interspeech 2021 Deep Noise Suppression (DNS) challenge. Evaluations on some other test sets also show the efficacy of our model.
\end{abstract}

\section{Introduction}
The widespread noise and reverberation may seriously degrade the performance of automatic speech recognition (ASR) systems and decrease speech intelligibility in communication. Speech enhancement aims at separating clean speech from background interference for higher speech intelligibility and perceptual quality. Despite the rapid progress of DNN-based speech enhancement recently, its performance in real applications still faces the challenges such as low signal-to-noise ratio (SNR), high reverberation and far-field pickup. The Interspeech 2021 deep noise suppression (DNS) challenge \cite{2021arXiv210101902R} is organized to foster more competitive speech enhancement system in adverse environments, and training datasets and evaluation metrics are provided for such purpose.

As a data-driven supervised learning approach, DNN-based speech enhancement can be mainly categorized into time-frequency domain \cite{6932438,8369155,7178061} and time domain \cite{8707065,8331910,2018arXiv180603185S} methods. The time-frequency (T-F) domain methods aim to extract the acoustic features (e.g., complex spectrum or logarithmic power spectrum) of clean speech from the features of noisy speech. Common training targets include ideal ratio mask (IRM) \cite{Hummersone} and target magnitude spectrum (TMS) \cite{8369155}, etc. The phase spectrum is also considered to benefit the speech quality \cite{Paliwal}. However, it is difficult to estimate phase spectrum directly because of its unstructured characteristic. Phase-sensitive mask (PSM) \cite{7178061} was proposed to exploit phase information for speech enhancement. More recent methods, such as PHASEN \cite{Yin}, make use of the inter-connection between the magnitude and phase spectrum for better phase estimation. Some other methods retrieve phase implicitly by optimizing the real and imaginary parts of the complex spectrum \cite{8910352} or estimating complex ratio mask (CRM) \cite{7364200}. Since complex-valued weights are suitable for modeling the inherent information of the spectrum, complex-valued neural networks \cite{2020arXiv200800264H} have also been used for speech enhancement.

The time domain methods directly estimate the clean speech waveforms through end-to-end training, circumventing the trouble of estimating phase information in the T-F domain. As a typical method in the time domain, Conv-Tasnet \cite{8707065} utilizes a 1-D convolution neural network (Conv-1D) \cite{Lea} as an encoder to convert time-domain waveform into effective representations for effective clean speech estimation, and then converts the representations back to waveform by a transposed convolutional layer called decoder. Time domain methods suffer from the difficulty of modeling extremely long sequences so that very deep convolutional layers like wave-u-net \cite{2018arXiv180603185S} have to be utilized for feature compression. Conventional recurrent neural networks (RNNs) are also not effective for modeling such long sequences. Dual-path recurrent neural network (DPRNN) \cite{9054266} was proposed to address this problem, in which the long sequential features are split into smaller chunks and processed by intra-chunk and inter-chunk RNNs iteratively, reducing the length of the sequence to be processed for every RNN. 

The intra-chunk operation in DPRNN aims at modeling the signal feature within a frame, which is also applicable in the frequency domain with the potential benefit of making full use of the harmonic spectral structure of speech. Thus it is reasonable to implement similar network in the T-F domain. When designing models for real-time speech enhancement, it is impractical to apply a convolutional neural network (CNN) with too many layers or non-causal structures like bidirectional long short-term memory (BiLSTM) \cite{Yin}. Recently a network structure called convolution recurrent network (CRN) \cite{8462155} is proposed. Taking advantage of both CNNs and RNNs, CRN can not only capture the local patterns of the spectrogram, but also model the dependence between consecutive frames. In this paper, we combine DPRNN and CRN in the T-F domain. On the basis of CRN, a new model called  \emph{dual-path convolution recurrent network} (DPCRN) is proposed. Similar to the DPRNN in the time domain, the DPCRN also uses two kinds of RNNs. The intra-chunk RNN is used to model the spectrum of a single time frame, while inter-chunk RNN is used to model the variation of the spectrum over time. The features compressed by convolutional layers are fed into DPRNN module for further processing, followed by a decoder composed of transposed convolutional layers. The CRM is output from the last transposed convolutional layer. We evaluate the DPCRN on the Interspeech 2021 DNS challenge dataset. Experimental results show that the DPCRN outperforms the baseline models, including NSNet2 \cite{9054254}, DTLN \cite{2020arXiv200507551W} and DCCRN \cite{2020arXiv200800264H}. On simulated test datasets, our model achieves competitive results as baseline models and show better performance in the case of low SNR. With only 0.8M parameters, our model achieves an overall MOS of 3.57 according to the ITU-T P.835 \cite{2020arXiv201013200N} subjective evaluation on DNS challenge blind test set, and reaches the third place in the wide band scenario track.
\section{Dual-Path Convolution Recurrent Network}
\subsection{Problem formulation}

In the time domain, the observed noisy speech can be formulated as \(x(t)=s(t)+n(t)\), where \(x(t)\), \(s(t)\) and \(n(t)\) refer to the noisy, the clean and the noise signals, respectively. The formula can be transformed into time-frequency domain by the short-time Fourier transform (STFT) as:
\begin{equation}
X(t,f)=S(t,f)+N(t,f){\rm{,}}\label{eq1}
\end{equation}
where \(X(t,f)\), \(S(t,f)\) and \(N(t,f)\) represent the time-frequency bin of the noisy, the clean and the noise speech spectrogram, respectively, at time frame \(t\) and frequency index \(f\). In order to recover clean speech from the mixture, a common way is to estimate a mask \(M(t,f)\) and multiply it by the noisy speech \(X(t,f)\) \cite{8369155}. For phase retrieval, we can separately estimate masks for magnitude and phase spectrogram or the real and imaginary parts of the complex spectrogram \cite{2020arXiv200800264H}. Another method is to directly estimate the complex ratio mask (CRM) \cite{7364200} which is denoted as \(M(t,f)=M_r(t,f)+iM_i(t,f)\), where \(M_r(t,f)\) and \(M_i(t,f)\) represent the real and imaginary parts of the mask. Then the denoising process can be expressed as the complex product of mask and noisy speech in the form of:
\begin{equation}
	\widetilde{S}(t,f)=X(t,f)\odot M(t,f),\label{eq2}
\end{equation}
where \(\odot\) denotes element-wise multiplication and \(\widetilde{S}(t,f)\) is the enhanced speech. Instead of estimating the mask directly, applying the signal approximation (SA) \cite{7032183} usually leads to better optimization. SA minimizes the difference between the enhanced speech and clean speech with the loss function described as \(\mathcal{L}=Loss(\widetilde{S}(t,f),S(t,f))\).

\subsection{Model architecture}

Dual-path RNN (DPRNN), originally proposed in \cite{9054266}, achieves state-of-the-art (SOTA) performance in single-channel speech separation task in the time domain. In this model, the speech waveform is converted into effective representations by an encoder which consists of Conv-1D layer. The separation is then performed by passing the encoder features to a well-designed DNN. For better performance, smaller kernel size of the Conv-1D is usually utilized, resulting in extremely long feature chunks. Conventional RNNs have trouble modeling such long sequences. In DPRNN, a long sequence is divided into overlapping chunks and processed by intra-chunk and inter-chunk RNNs for better optimization. Recently, the DPRNN has also been combined with self-attention mechanism for time-domain speech enhancement \cite{2020arXiv201012713P}.

\begin{figure*}[tbh!]
	\centering
	\vspace{-0.4cm} 
	
	\begin{minipage}[tbh!]{0.8\linewidth}
		\centering
		\includegraphics[width=\linewidth,height=0.10\textheight]{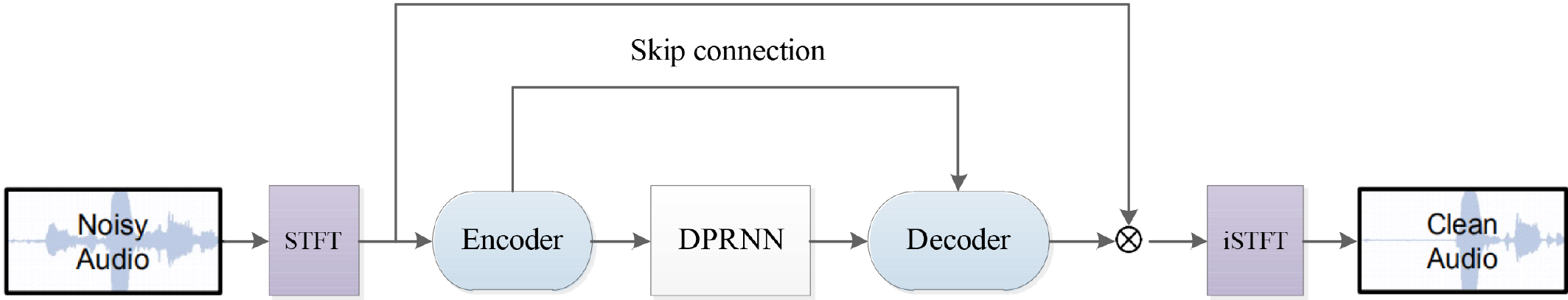}
	\end{minipage}%
	\label{fig:subfig:DPCRN}
	
	(a)
	
	\begin{minipage}[tbh!]{0.8\linewidth}
		\centering
		\includegraphics[width=\linewidth,height=0.11\textheight]{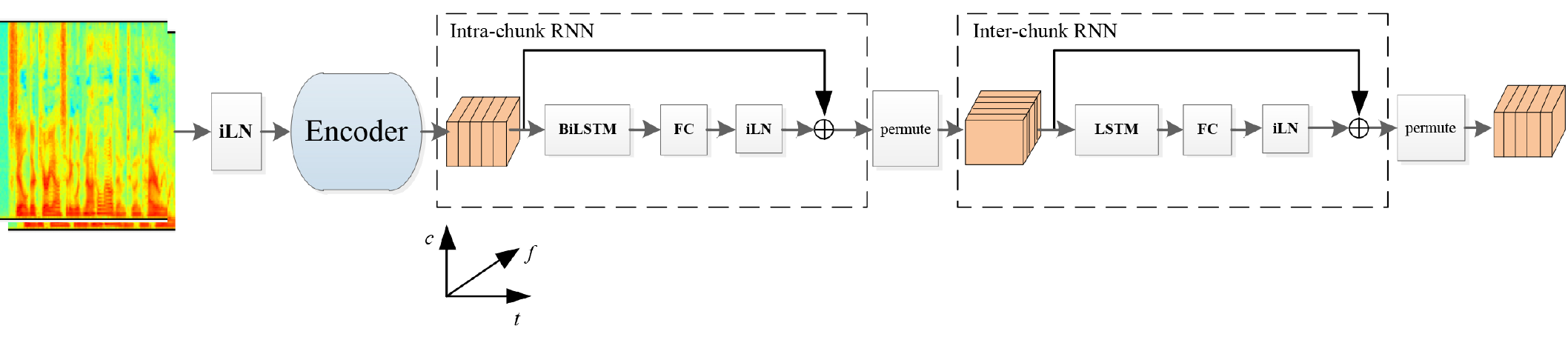}
	\end{minipage}%
	\label{fig:subfig:DPRNN}
	
	(b)
	\caption{(a) Proposed DPCRN model and (b) the diagram of the DPRNN module. “f”, “t” and “c” represent frequency, time and channel axes, respectively.}
	\label{fig:figure1}
	\vspace{-0.5cm}
\end{figure*}

The intra-chunk operation in DPRNN is also applicable in the frequency domain with the potential benefit of making full use of the spectral structure of speech. By combining DPRNN and CRN, it is possible to obtain a well-behaved model in the T-F domain. Similar to the original DPRNN, our model consists of an encoder, a dual-path RNN module and a decoder, as shown in Figure~\ref{fig:figure1}(a). The structure of the encoder and decoder is similar to CRN \cite{8462155}. We send the real and imaginary parts of complex spectrogram of the noisy signal into the encoder as two streams. The encoder uses the 2-D convolutional (Conv-2D) layers to extract local patterns from noisy spectrogram and reduce the feature resolution. The decoder uses transposed convolutional layers to restore low-resolution features to the original size, forming a symmetric structure with the encoder. There are skip connections between the encoder and the decoder to pass the detailed information. Every convolutional layer is followed by a batch normalization and a PReLU function \cite{7410480}. We replace the RNN part of the CRN with the DPRNN module, as depicted in Figure~\ref{fig:figure1}(b). Different from the original DPRNN, we regard the frames in STFT as the chunks for DPRNN processing. Instead of learning the dependence in the time domain, the intra-chunk RNNs are applied to model the spectral patterns in a single frame. We believe that modeling the dependence of frequency is beneficial to speech enhancement due to the harmonic structures of speech. RNNs can overcome the shortcoming of limited receptive field of CNNs and capture long-term harmonic correlation. As for the inter-chunk RNNs, we use LSTM to model the time dependence of a certain frequency, so that a strict causality can be guaranteed. These LSTMs are computed in parallel. BiLSTM is used for intra-chunk modeling, which will not influence the causality of the whole system. The LSTM and BiLSTM are followed by a fully-connected layer (FC) and a layer normalization (LN) \cite{2016arXiv160706450L}. A residual connection \cite{7780459} is then applied between the input of RNN and the output of LN to further mitigate the gradient vanishing problem. 

Instead of the common LN, we use instant layer normalization (iLN) \cite{2020arXiv200507551W} in our model, where all frames calculate statistics independently on frequency axis \(f\) and channel axis \(c\), and share the same trainable parameters. Denote \(\mathbf{F}_t\in\mathbb{R}^{N\times K}\) as the feature matrix of the \(t\)-th frame, \(N\) and \(K\) the feature dimensions of \(f\) and \(c\), \(\hat{E}\) and \(\hat{D}\) the mean and variance operator, \(\bm{\gamma}\) and \(\bm{\beta}\in\mathbb{R}^{N\times K}\) trainable parameters, and \(\varepsilon\) a regularization parameter, then the iLN for the feature at time index \(t\) is defined as:
\begin{equation}
	iLN\left(\mathbf{F}_t\right)=\frac{\mathbf{F}_t-\hat{E}\left[\mathbf{F}_t\right]}{\sqrt{\hat{D}\left[\mathbf{F}_t\right]+\varepsilon}}\odot\bm{\gamma}+\bm{\beta},\label{eq3}
\end{equation}
where
\begin{equation}
	\hat{E}\left[\mathbf{F}_t\right]=\frac{1}{NK}\sum_{f=1}^{N}\sum_{c=1}^{K}{\mathbf{F}_t\left[f,c\right]}\label{eq5}
\end{equation}
and
\begin{equation}
	\hat{D}\left[\mathbf{F}_t\right]=\frac{1}{NK}\sum_{f=1}^{N}\sum_{c=1}^{K}\left(\mathbf{F}_t\left[f,c\right]-\hat{E}\left[\mathbf{F}_t\right]\right)^2.\label{eq5}
\end{equation}
To reduce the sensitivity of the model output to the energy of the input signal, we also apply the iLN on the input spectrogram.
\subsection{Learning Target and Loss function}
In our experiments, the learning target of DPCRN is CRM. The real and imaginary parts of the CRM are outputs from the decoder as two streams. In training stage, the learning target is optimized by signal approximation (SA). Multiplying the spectrogram of noisy speech \(X=X_r+iX_i\) with the estimated mask \(M=M_r+iM_i\), we get enhanced spectrogram in the form of:
\begin{equation}
	\widetilde{S}=X_rM_r-X_iM_i+i(X_rM_i+X_iM_r),\label{eq6}
\end{equation}
which is converted back to the waveform using inverse STFT (iSTFT):
\begin{equation}
	\widetilde{s}=\mathrm{iSTFT}(\widetilde{S}).\label{eq7}
\end{equation}
We use two loss function in the experiments for comparation. The first loss function \(f\) is the negative signal-to-noise ratio (SNR) \cite{8937253} defined as:
\begin{equation}
	f(s,\widetilde{s})=-10\mathrm{log}_{\mathrm{10}}(\frac{\sum_{t}{s(t)^2}}{\sum_{t}(s(t)-\widetilde{s}(t))^2}).\label{eq8}
\end{equation}
Compared with commonly used scale invariant SNR (SI-SNR), it can constrain the amplitude of the output and avoid level offset between input and output, which is important for real-time processing. Taking the spectrogram quality into consideration, we add the mean square error (MSE) of the spectrogram to the negative SNR and get the second loss function, which is defined as:
\begin{equation}
	\begin{split}
	\mathcal{L}_{\mathrm{MSE}}=&f(s,\widetilde{s})+\mathrm{log}(\mathrm{MSE}(S_r,\widetilde{S}_r)+\mathrm{MSE}(S_i,\widetilde{S}_i)\\
	&+\mathrm{MSE}(|S|,|\widetilde{S}|)).\label{eq9}
	\end{split}
\end{equation}
The added MSE loss consists of three parts, which respectively measure the difference of real part, imaginary part and magnitude between the estimated spectrogram and the true one. We take the logarithm of the MSE loss to ensure that it is of the same order of magnitude as the negative SNR.
\section{Experiments}

\subsection{Datasets}
We trained the DPCRN on the Interspeech 2021 DNS challenge dataset. 60000 clips of reverberant speech (about 500 h) were generated, with 55000 clips for training and 5000 for validation. The noise clips were mainly generated from Audioset \cite{7952261}, DEMAND \cite{demand} and Freesound\footnote{https://freesound.org/}. In training stage, we randomly split the waves into 5-second segments and convolved them with room impulse responses (RIRs) randomly-selected from openSLR26 and openSLR28 \cite{7953152}. Then the noisy speech was generated by mixing reverberant speech and noise. The SNR range of the mixture is set between -5 and 5 dB.

In order to test the performance under various unknown noise, we also used the test set from WSJ-0 \cite{wsj0} as the test speech. It contains 651 utterances from 8 speakers. There are two noise datasets used for test; one is the music data from MUSAN \cite{2015arXiv151008484S}, the other is babble, factory1 and f16 from NOISEX92 \cite{NOISEX-92}. The SNR range of the test noisy speech is the same as the training set. We also evaluated the model on the development test set and blind test set provided by DNS challenge. All the audio used is sampled at 16kHz.

\subsection{Parameter setup}
In our model, the window length and hop size are 25ms and 12.5ms respectively, resulting in a total latency of 37.5ms, which satisfies the DNS challenge requirement. The FFT length is 400 and the sine window is applied before FFT and overlap-add. The 201-dimensional complex spectrum is fed into the model. The channel number of the convolutional layers in the encoder is \{32,32,32,64,128\}. The kernel size and the stride are respectively set to \{(5,2),(3,2),(3,2),(3,2),(3,2)\} and \{(2,1),(2,1),(1,1),(1,1),(1,1)\} in frequency and time dimension. All the Conv-2D and transposed Conv-2D layers are causally computed. We use two DPRNN modules, each of them has RNNs with a hidden dimension of 128. The total model parameters is about 0.8M. There are three models for comparison in the following experiments. The models with loss functions shown in equation (8) and (9) are called DPCRN-1 and DPCRN-2, respectively. In the third model, called DPCRN-3, we set the stride to \{(2,1),(2,1),(2,1),(1,1),(1,1)\}, reducing frequency resolution of the feature fed into the DPRNN by half. At the same time, we double the hidden dimension of the intra-chunk RNN for the same computational complexity. The loss function of DPCRN-3 is the same as DPCRN-1. 

The models are trained by Adam optimizer \cite{2014arXiv1412.6980K} and the batch size is 8. The initial learning rate is 1e-3 and it will be halved if the loss on the validation set does not improve for five consecutive epochs. Early stopping is also applied in training if the loss on the validation set does not improve for ten epochs. TensorFlow is employed for model implementation and a Nvidia GeForce GTX 1080Ti is used for training.
\begin{table*}[!htbp]
	\vspace{-0.8cm} 
	\setlength{\abovecaptionskip}{2pt}
	\caption{Experimental results on WSJ0-MUSAN test set. \textbf{BLOD} indicates the best result for each case.}
	\label{tab:table1}
	\setlength{\tabcolsep}{2mm}
	\centering
	\begin{tabular}{ccccccccccccccc}
		\toprule
		Metrics & \multicolumn{4}{c}{PESQ}                           & \textbf{} & \multicolumn{4}{c}{{STOI (in \%)}}                        & \multicolumn{1}{l}{\textbf{}} & \multicolumn{4}{c}{{SDR (in dB)}}                        \\
		\cline{2-5} \cline{7-10}  \cline{12-15}
		SNR(dB)          & -5            & 0             & 5             & Avg.          &           & -5            & 0              & 5              & Avg.           &                               & -5            & 0              & 5              & Avg.           \\
		\midrule
		Noisy            & 1.67          & 2.01          & 2.35          & 2.01          &           & 71.10          & 81.44          & 89.63          & 80.72          &                               & -4.91         & 0.04           & 5.02           & 0.05           \\
		DTLN             & 2.23          & 2.62          & 2.93          & 2.59          &           & 82.53         & 90.52          & 94.98          & 89.34          &                               & 5.74          & 10.01          & 13.77          & 9.85           \\
		DCCRN            & 2.42          & 2.87          & \textbf{3.23} & 2.84          &           & 86.40          & \textbf{93.08} & \textbf{96.63} & \textbf{92.04} &                               & 6.17          & 11.08          & \textbf{15.05} & 10.76          \\
		DPCRN-1          & 2.51          & 2.85          & 3.14          & 2.83          &           & 86.78         & 92.70           & 95.97          & 91.82          &                               & \textbf{7.48} & \textbf{11.17} & 14.46          & \textbf{11.04} \\
		DPCRN-2          & \textbf{2.52} & \textbf{2.87} & 3.15          & \textbf{2.85} &           & \textbf{87.10} & 92.93          & 96.04          & 92.02          &                               & 7.40           & 11.14          & 14.44          & 10.99          \\
		DPCRN-3          & 2.33          & 2.68          & 2.97          & 2.66          &           & 84.26         & 91.02          & 94.96          & 90.08          &                               & 6.12          & 10.02          & 13.49          & 9.88           \\    
		\bottomrule
	\end{tabular}
\end{table*}
\begin{table*}[!htbp]
	\vspace{-0.3cm} 
	\setlength{\abovecaptionskip}{2pt}
	\caption{Experimental results on WSJ0-NOISEX92 test set. \textbf{BLOD} indicates the best result for each case.}
	\label{tab:table2}
	\setlength{\tabcolsep}{2.08mm} 
	\centering
	\begin{tabular}{ccccccccccccccc}
		\toprule
		Metrics & \multicolumn{4}{c}{PESQ}                           & \textbf{} & \multicolumn{4}{c}{{STOI (in \%)}}                        & \multicolumn{1}{l}{\textbf{}} & \multicolumn{4}{c}{{SDR (in dB)}}                        \\
		\cline{2-5} \cline{7-10}  \cline{12-15}
		SNR(dB)          & -5            & 0             & 5             & Avg.          &           & -5            & 0              & 5              & Avg.           &                               & -5            & 0              & 5              & Avg.           \\
		\midrule
		Noisy            & 1.43          & 1.71          & 2.05          & 1.72          &           & 62.79         & 75.47          & 86.23          & 74.94          &                               & -4.93         & 0.04           & 5.02           & 0.09           \\
		DTLN             & 1.91          & 2.34          & 2.67          & 2.31          &           & 72.72         & 85.90          & 92.68          & 83.91          &                               & 4.15          & 8.55          & 12.36          & 8.42          \\
		DCCRN            & 1.85          & 2.34          & 2.78          & 2.32          &           & 74.51         & 87.87          & \textbf{94.38} & 85.59          &                               & 2.79          & 8.22          & 12.60          & 7.87          \\
		DPCRN-1          & 2.04          & 2.46          & 2.80          & 2.43          &           & \textbf{77.00}& 88.20          & 93.80          & 86.33          &                               & \textbf{5.36} & \textbf{9.45} & \textbf{12.96}  & \textbf{9.25} \\
		DPCRN-2          & \textbf{2.05} & \textbf{2.49} & \textbf{2.83} & \textbf{2.46} &           & 76.98         & \textbf{88.28} & 93.88          & \textbf{86.38} &                               & 5.22          & 9.37          & 12.90          & 9.16         \\
		DPCRN-3          & 1.90          & 2.31          & 2.66          & 2.29          &           & 74.86         & 86.61          & 92.88          & 84.78          &                               & 4.51          & 8.70          & 12.28          & 8.50           \\    
		\bottomrule
	\end{tabular}
	\vspace{-0.5cm}
\end{table*}
\setlength{\belowcaptionskip}{-5pt}
\subsection{Baselines and evaluation metrics}
We compare our model with the top-ranking models in Interspeech 2020 DNS challenge on the first test set, including DTLN \cite{2020arXiv200507551W} and DCCRN \cite{2020arXiv200800264H}. DTLN combines the STFT and a learnable transformation with only 1M parameters. DCCRN uses complex-valued convolution neural networks and got the first place in the real-time track. The baseline model NSNet2 \cite{9054254} provided by Interspeech 2021 DNS challenge is also compared on the DNS test set.

On the simulated WSJ0 test set, we use three objective evaluation metrics: perceptual evaluation of speech quality (PESQ) \cite{941023}, shorter-time objective intelligibility (STOI) \cite{STOI} and signal to distortion Ratio (SDR) \cite{1643671}. On the DNS challenge development test set, we use DNSMOS \cite{2020arXiv201015258R} for evaluation, which is a DNN-based non-intrusive speech quality evaluation metric. A subjective evaluation according to ITU-T P.835 \cite{2020arXiv201013200N} was also applied on the DNS blind test set as the final result.

\begin{table}[]
	\vspace{0.2cm}
	\setlength{\abovecaptionskip}{2pt}
	\caption{DNSMOS on DNS challenge development test set.}
	\label{tab:table3}
	\centering
	\begin{tabular}{ccccl}
		\toprule
		Model   & Para. (M) & look-ahead (ms) & DNSMOS         &  \\
		\midrule
		Noisy   & -         & -               & 2.899          &  \\
		NSNet2  & 2.8       & 0               & 3.243          &  \\
		DTLN    & 1.0         & 0               & 3.226          &  \\
		DCCRN   & 3.7       & 37.5            & 3.373          &  \\
		DPCRN-1 & 0.8       & 0               & 3.454          &  \\
		DPCRN-2 & 0.8       & 0               & \textbf{3.472} &  \\
		\bottomrule
	\end{tabular}
	\vspace{-0.5cm}
\end{table}
\begin{table}[th]
	\setlength{\abovecaptionskip}{2pt}
	\caption{Performance on DNS challenge blind test set.}
	\label{tab:table4}
	\setlength{\tabcolsep}{2.81mm}
	\centering
	\begin{tabular}{cp{30pt}p{50pt}p{30pt}l}
		\toprule
		\makecell[c]{Model} & \makecell[c]{Speech\\MOS} & \makecell[c]{Background\\Noise MOS} & \makecell[c]{Overall\\MOS} &  \\
		\midrule
		Noisy          & \makecell[c]{3.89}                & \makecell[c]{2.60}                           & \makecell[c]{2.77}                 &  \\
		NSNet2         & \makecell[c]{3.35}                & \makecell[c]{3.88}                          & \makecell[c]{3.07}                 &  \\
		DPCRN-2        & \makecell[c]{\textbf{3.76}}       & \makecell[c]{\textbf{4.34}}                 & \makecell[c]{\textbf{3.57}}        &  \\
		\bottomrule
	\end{tabular}
\end{table}
\vspace{-0.5cm} 
\subsection{Results and analysis}
The performance on simulated WSJ0-MUSAN test set is presented in Table~\ref{tab:table1}. It can be seen that when the SNR is greater than or equal to 0 dB, the performance of DPCRN-1 is slightly weaker than DCCRN, but better than DTLN. It should be noted that DPCRN-1 performs better than DCCRN at lower SNR. Table~\ref{tab:table2} shows the results on WSJ0-NOISEX92 test set. Under more disruptive noise from NOISEX92, the DPCRN-1 exceeds the baseline models in terms of all three metrics, demonstrating the benefit of the DPRNN module for spectrogram modeling. On both datasets, DPCRN-2 has better performance than DPCRN-1 in terms of PESQ and STOI but its SDR is slightly worse, indicating that including the time-frequency MSE in the loss function has benefit to speech quality. DPCRN-3 has more parameters than DPCRN-1, but its performance is worse, which indicates that reducing the frequency resolution of the features fed into the DPRNN is detrimental to the system. RNN faces the difficulty of parallel computation, which is a challenge to real-time processing. In our submitted model, we set the frequency dimension of the feature to 50 to guarantee a decent frequency resolution while meeting the real-time requirements of DNS challenge.

In Table~\ref{tab:table3}, we compare DPCRN with baseline models on the DNS challenge development test set. “Para.” and “look-ahead” in the table respectively represent the parameter amount of the model and the length of used future information. With about 0.8M parameters and without any future information, our models perform better than the baseline models in terms of DNSMOS, among which the DPCRN-2 is the best one. Therefore, we choose this model for the wide band scenario track. The final P.835 MOS on the DNS challenge blind test set are shown in Table~\ref{tab:table4}, which gives speech, background noise and overall MOS. With a little speech quality deterioration, the submitted model achieves an overall MOS of 3.57 and ranks the third in the wide band scenario track. The computation complexity of our model is about 7.45 Giga floating-point operations per second (GFLOPS) and the one frame processing time of the TensorFlow implementation is 8.9 ms on a quad-core Intel i5-6300HQ.
\section{Conclusions}
Inspired by the successful application of DPRNN and CRN, we propose a deep learning-based speech enhancement model in the time-frequency domain, named as DPCRN. It combines the local pattern modeling capability of CNN and the long-term modeling capability of DPRNN. Compared with CRN, DPCRN demonstrates the benefit of RNN for spectrum modeling. With only 0.8M parameters, our model achieves competitive results on various unknown noise datasets. In the future, we will try to reduce the computational complexity of the model for wider band spectrum processing.

\section{Acknowledgement}
The National Science Foundation of China supported this work with grant number 11874219.

\bibliographystyle{IEEEtran}

\bibliography{mybib}

\begin{thebibliography}{10}
\providecommand{\url}[1]{#1}
\csname url@samestyle\endcsname
\providecommand{\newblock}{\relax}
\providecommand{\bibinfo}[2]{#2}
\providecommand{\BIBentrySTDinterwordspacing}{\spaceskip=0pt\relax}
\providecommand{\BIBentryALTinterwordstretchfactor}{4}
\providecommand{\BIBentryALTinterwordspacing}{\spaceskip=\fontdimen2\font plus
\BIBentryALTinterwordstretchfactor\fontdimen3\font minus
  \fontdimen4\font\relax}
\providecommand{\BIBforeignlanguage}[2]{{%
\expandafter\ifx\csname l@#1\endcsname\relax
\typeout{** WARNING: IEEEtran.bst: No hyphenation pattern has been}%
\typeout{** loaded for the language `#1'. Using the pattern for}%
\typeout{** the default language instead.}%
\else
\language=\csname l@#1\endcsname
\fi
#2}}
\providecommand{\BIBdecl}{\relax}
\BIBdecl

\bibitem{2021arXiv210101902R}
C.~K.~A. {Reddy}, H.~{Dubey}, K.~{Koishida}, A.~{Nair}, V.~{Gopal},
  R.~{Cutler}, S.~{Braun}, H.~{Gamper}, R.~{Aichner}, and S.~{Srinivasan},
  ``{Interspeech 2021 Deep Noise Suppression Challenge},'' \emph{arXiv
  e-prints}, p. arXiv:2101.01902, 2021.

\bibitem{6932438}
Y.~{Xu}, J.~{Du}, L.~{Dai}, and C.~{Lee}, ``A regression approach to speech
  enhancement based on deep neural networks,'' \emph{IEEE/ACM Transactions on
  Audio, Speech, and Language Processing}, vol.~23, no.~1, pp. 7--19, 2015.

\bibitem{8369155}
D.~{Wang} and J.~{Chen}, ``Supervised speech separation based on deep learning:
  An overview,'' \emph{IEEE/ACM Transactions on Audio, Speech, and Language
  Processing}, vol.~26, no.~10, pp. 1702--1726, 2018.

\bibitem{7178061}
H.~{Erdogan}, J.~R. {Hershey}, S.~{Watanabe}, and J.~{Le Roux},
  ``Phase-sensitive and recognition-boosted speech separation using deep
  recurrent neural networks,'' in \emph{2015 IEEE International Conference on
  Acoustics, Speech and Signal Processing (ICASSP)}, 2015, pp. 708--712.

\bibitem{8707065}
Y.~{Luo} and N.~{Mesgarani}, ``Conv-tasnet: Surpassing ideal time–frequency
  magnitude masking for speech separation,'' \emph{IEEE/ACM Transactions on
  Audio, Speech, and Language Processing}, vol.~27, no.~8, pp. 1256--1266,
  2019.

\bibitem{8331910}
S.~{Fu}, T.~{Wang}, Y.~{Tsao}, X.~{Lu}, and H.~{Kawai}, ``End-to-end waveform
  utterance enhancement for direct evaluation metrics optimization by fully
  convolutional neural networks,'' \emph{IEEE/ACM Transactions on Audio,
  Speech, and Language Processing}, vol.~26, no.~9, pp. 1570--1584, 2018.

\bibitem{2018arXiv180603185S}
D.~{Stoller}, S.~{Ewert}, and S.~{Dixon}, ``{Wave-U-Net: A Multi-Scale Neural
  Network for End-to-End Audio Source Separation},'' \emph{arXiv e-prints}, p.
  arXiv:1806.03185, 2018.

\bibitem{Hummersone}
C.~Hummersone, T.~Stokes, and T.~Brookes, \emph{On the Ideal Ratio Mask as the
  Goal of Computational Auditory Scene Analysis}, 2014, pp. 349--368.

\bibitem{Paliwal}
K.~Paliwal, K.~Wojcicki, and B.~Shannon, ``The importance of phase in speech
  enhancement,'' \emph{Speech Communication}, vol.~53, pp. 465--494, 2011.

\bibitem{Yin}
D.~Yin, C.~Luo, Z.~Xiong, and W.~Zeng, ``Phasen: A phase-and-harmonics-aware
  speech enhancement network,'' \emph{Proceedings of the AAAI Conference on
  Artificial Intelligence}, vol.~34, pp. 9458--9465, 2020.

\bibitem{8910352}
K.~{Tan} and D.~{Wang}, ``Learning complex spectral mapping with gated
  convolutional recurrent networks for monaural speech enhancement,''
  \emph{IEEE/ACM Transactions on Audio, Speech, and Language Processing},
  vol.~28, pp. 380--390, 2020.

\bibitem{7364200}
D.~S. {Williamson}, Y.~{Wang}, and D.~{Wang}, ``Complex ratio masking for
  monaural speech separation,'' \emph{IEEE/ACM Transactions on Audio, Speech,
  and Language Processing}, vol.~24, no.~3, pp. 483--492, 2016.

\bibitem{2020arXiv200800264H}
Y.~{Hu}, Y.~{Liu}, S.~{Lv}, M.~{Xing}, S.~{Zhang}, Y.~{Fu}, J.~{Wu},
  B.~{Zhang}, and L.~{Xie}, ``{DCCRN: Deep Complex Convolution Recurrent
  Network for Phase-Aware Speech Enhancement},'' \emph{arXiv e-prints}, p.
  arXiv:2008.00264, 2020.

\bibitem{Lea}
C.~Lea, R.~Vidal, A.~Reiter, and G.~Hager, ``Temporal convolutional networks: A
  unified approach to action segmentation,'' 2016.

\bibitem{9054266}
Y.~{Luo}, Z.~{Chen}, and T.~{Yoshioka}, ``Dual-path rnn: Efficient long
  sequence modeling for time-domain single-channel speech separation,'' in
  \emph{ICASSP 2020 - 2020 IEEE International Conference on Acoustics, Speech
  and Signal Processing (ICASSP)}, 2020, pp. 46--50.

\bibitem{8462155}
H.~{Zhao}, S.~{Zarar}, I.~{Tashev}, and C.~{Lee}, ``Convolutional-recurrent
  neural networks for speech enhancement,'' in \emph{2018 IEEE International
  Conference on Acoustics, Speech and Signal Processing (ICASSP)}, 2018, pp.
  2401--2405.

\bibitem{9054254}
Y.~{Xia}, S.~{Braun}, C.~K.~A. {Reddy}, H.~{Dubey}, R.~{Cutler}, and
  I.~{Tashev}, ``Weighted speech distortion losses for neural-network-based
  real-time speech enhancement,'' in \emph{ICASSP 2020 - 2020 IEEE
  International Conference on Acoustics, Speech and Signal Processing
  (ICASSP)}, 2020, pp. 871--875.

\bibitem{2020arXiv200507551W}
N.~L. {Westhausen} and B.~T. {Meyer}, ``{Dual-Signal Transformation LSTM
  Network for Real-Time Noise Suppression},'' \emph{arXiv e-prints}, p.
  arXiv:2005.07551, 2020.

\bibitem{2020arXiv201013200N}
B.~{Naderi} and R.~{Cutler}, ``{A Crowdsourcing Extension of the ITU-T
  Recommendation P.835 with Validation},'' \emph{arXiv e-prints}, p.
  arXiv:2010.13200, 2020.

\bibitem{7032183}
F.~{Weninger}, J.~R. {Hershey}, J.~{Le Roux}, and B.~{Schuller},
  ``Discriminatively trained recurrent neural networks for single-channel
  speech separation,'' in \emph{2014 IEEE Global Conference on Signal and
  Information Processing (GlobalSIP)}, 2014, pp. 577--581.

\bibitem{2020arXiv201012713P}
A.~{Pandey} and D.~{Wang}, ``{Dual-path Self-Attention RNN for Real-Time Speech
  Enhancement},'' \emph{arXiv e-prints}, p. arXiv:2010.12713, 2020.

\bibitem{7410480}
K.~{He}, X.~{Zhang}, S.~{Ren}, and J.~{Sun}, ``Delving deep into rectifiers:
  Surpassing human-level performance on imagenet classification,'' in
  \emph{2015 IEEE International Conference on Computer Vision (ICCV)}, 2015,
  pp. 1026--1034.

\bibitem{2016arXiv160706450L}
J.~{Lei Ba}, J.~R. {Kiros}, and G.~E. {Hinton}, ``{Layer Normalization},''
  \emph{arXiv e-prints}, p. arXiv:1607.06450, Jul. 2016.

\bibitem{7780459}
K.~{He}, X.~{Zhang}, S.~{Ren}, and J.~{Sun}, ``Deep residual learning for image
  recognition,'' in \emph{2016 IEEE Conference on Computer Vision and Pattern
  Recognition (CVPR)}, 2016, pp. 770--778.

\bibitem{8937253}
I.~{Kavalerov}, S.~{Wisdom}, H.~{Erdogan}, B.~{Patton}, K.~{Wilson}, J.~{Le
  Roux}, and J.~R. {Hershey}, ``Universal sound separation,'' in \emph{2019
  IEEE Workshop on Applications of Signal Processing to Audio and Acoustics
  (WASPAA)}, 2019, pp. 175--179.

\bibitem{7952261}
J.~F. {Gemmeke}, D.~P.~W. {Ellis}, D.~{Freedman}, A.~{Jansen}, W.~{Lawrence},
  R.~C. {Moore}, M.~{Plakal}, and M.~{Ritter}, ``Audio set: An ontology and
  human-labeled dataset for audio events,'' in \emph{2017 IEEE International
  Conference on Acoustics, Speech and Signal Processing (ICASSP)}, 2017, pp.
  776--780.

\bibitem{demand}
J.~Thiemann, N.~Ito, and E.~Vincent, ``The diverse environments multi-channel
  acoustic noise database (demand): A database of multichannel environmental
  noise recordings,'' \emph{The Journal of the Acoustical Society of America},
  vol. 133, p. 3591, 2013.

\bibitem{7953152}
T.~{Ko}, V.~{Peddinti}, D.~{Povey}, M.~L. {Seltzer}, and S.~{Khudanpur}, ``A
  study on data augmentation of reverberant speech for robust speech
  recognition,'' in \emph{2017 IEEE International Conference on Acoustics,
  Speech and Signal Processing (ICASSP)}, 2017, pp. 5220--5224.

\bibitem{wsj0}
D.~Paul and J.~Baker, ``The design for the wall street journal-based csr
  corpus,'' 1992.

\bibitem{2015arXiv151008484S}
D.~{Snyder}, G.~{Chen}, and D.~{Povey}, ``{MUSAN: A Music, Speech, and Noise
  Corpus},'' \emph{arXiv e-prints}, p. arXiv:1510.08484, 2015.

\bibitem{NOISEX-92}
A.~Varga and H.~Steeneken, ``Assessment for automatic speech recognition: Ii.
  noisex-92: A database and an experiment to study the effect of additive noise
  on speech recognition systems,'' \emph{Speech Communication}, vol.~12, pp.
  247--251, 1993.

\bibitem{2014arXiv1412.6980K}
D.~P. {Kingma} and J.~{Ba}, ``{Adam: A Method for Stochastic Optimization},''
  \emph{arXiv e-prints}, p. arXiv:1412.6980, 2014.

\bibitem{941023}
A.~W. {Rix}, J.~G. {Beerends}, M.~P. {Hollier}, and A.~P. {Hekstra},
  ``Perceptual evaluation of speech quality (pesq)-a new method for speech
  quality assessment of telephone networks and codecs,'' in \emph{2001 IEEE
  International Conference on Acoustics, Speech, and Signal Processing.
  Proceedings (Cat. No.01CH37221)}, vol.~2, 2001, pp. 749--752 vol.2.

\bibitem{STOI}
C.~Taal, R.~Hendriks, R.~Heusdens, and J.~Jensen, ``A short-time objective
  intelligibility measure for time-frequency weighted noisy speech,'' 2010, pp.
  4214 -- 4217.

\bibitem{1643671}
E.~{Vincent}, R.~{Gribonval}, and C.~{Fevotte}, ``Performance measurement in
  blind audio source separation,'' \emph{IEEE Transactions on Audio, Speech,
  and Language Processing}, vol.~14, no.~4, pp. 1462--1469, 2006.

\bibitem{2020arXiv201015258R}
C.~K.~A. {Reddy}, V.~{Gopal}, and R.~{Cutler}, ``{DNSMOS: A Non-Intrusive
  Perceptual Objective Speech Quality metric to evaluate Noise Suppressors},''
  \emph{arXiv e-prints}, p. arXiv:2010.15258, 2020.

\end{thebibliography}

\end{document}